\documentclass{aa}  

\usepackage{graphicx}
\usepackage{txfonts}
\usepackage{placeins}           
                                
\usepackage[dvipsnames]{xcolor}
\usepackage{url}
\usepackage{amsmath}
\usepackage{afterpage}
\usepackage{hyperref}
\makeatletter
\newlength\wtparindent
\makeatother

\begin{document}

   \title{CMBolic: Symbolic emulators for the Cosmic Microwave Background}

   \subtitle{I. Lensing}

%
%
%

   \author{D. M. J. Vokrouhlicky\inst{1,2}\corrauth{vokrouhlicky@fzu.cz}       
        \and C. Skordis\inst{1,3}
        \and D.J. Bartlett\inst{3}
        \and H. Desmond\inst{4}
        \and P.G. Ferreira\inst{3}
        }

   \institute{CEICO, Institute of Physics of the Czech Academy of Sciences, Na Slovance 1999/2, 182 00, Prague, Czech Republic
   \and Institute of Theoretical Physics, Faculty of Mathematics and Physics, Charles University, V Holešovičkách 747/2, 180 00 Prague~8, Czech Republic
   \and Astrophysics, University of Oxford, Denys Wilkinson Building, Keble Road, Oxford OX1 3RH, United Kingdom
   \and Institute of Cosmology \& Gravitation, University of Portsmouth, Dennis Sciama Building, Portsmouth, PO1 3FX, UK}

   \date{}

 
  \abstract{
We present the first installment of CMBolic: a suite of symbolic cosmic microwave background (CMB) emulators. In this instance,
we emulate the CMB lensing potential power spectrum $C_\ell^{\phi\phi}$ for the widely used extended $\Lambda$CDM model which simultaneously includes massive neutrinos
and evolving dark energy modelled using the Chevallier-Polarski-Linder (CPL) parameterization. 
We achieve comparable precision to existing neural network emulators, with the added benefit of simpler handling as our emulators are analytic functions of the model parameters and multipole $\ell$. 
On independent validation spectra evaluated in the range \(2\leq \ell \leq 5500\), CMBolic achieves mean absolute fractional errors of \(0.27\%\) in the \(\Lambda\)CDM subspace and \(0.32\%\) across the 
full extended parameter space. This emulation error is well below even the most optimistic noise forecasts from CMB Stage 4 experiments. 
We apply CMBolic to cosmological parameter estimation with Bayesian inference using the lensing-only likelihoods from ACT DR6 and Planck.
We show excellent agreement between the posteriors obtained by CMBolic and the
Boltzmann code CLASS.
This demonstrates the practical use of CMBolic on cosmological parameter estimation, reducing the runtime from 2 weeks to under 3 minutes. 
 }


   \maketitle

\section{Introduction}

Observations of the cosmic microwave background (CMB) anisotropies, including measurements from Planck and ACT/AdvACT, have provided exceptionally precise measurements of the parameters defining the standard $\Lambda$CDM cosmological model \citep{aghanim2020planck,henderson2016advanced}.
In the coming years, next-generation CMB experiments, such as Simons Observatory (SO) \citep{ade2019simons} and its Large Aperture Telescope \citep{abitbol2025simons}, 
and a reduced version of the S4 project \citep{abazajian2019cmb}, will push data and constraints even further.

Computational cosmology with large datasets brings the need for fast inference. 
It is standard to use Einstein-Boltzmann (EB) solvers such as CLASS \citep{lesgourgues2011cosmic} or CAMB \citep{lewis2011camb} to calculate theoretical predictions for observables.
However, these can be slow, especially in extended cosmologies and with the precision  necessary for Stage 4 surveys. 
To run full Markov chain Monte Carlo (MCMC) chains and apply Bayesian inference, at least $10^5-10^6$ evaluations are needed for the current cosmological landscape. To mitigate computational constraints, two primary 
strategies have been adopted: acceleration of the theoretical computation of power spectra, or new Bayesian methods requiring fewer evaluations to converge. 
The latter would still benefit from faster model computation. 

Approaches to accelerating Bayesian inference rely on more efficient sampling methods \citep{nygaard2023connect}, or, for instance, on decreasing the number of evaluations needed,
 by using neural network (NN) accelerators \citep{to2023linna}, Gaussian process and PCA based approaches \citep{gunther2023uncertainty}.
From the side of emulation of cosmological observables, several emulators have become widely used, such as the Euclid emulator~\citep{Euclid:2020rfv}, and the Bacco Emulator suite for total matter power spectra, both in the linear \citep{bacco2021linear} and nonlinear case \citep{bacco2021nonlinear}, and CosmoPower
\citep{spurio2022cosmopower}, which emulates matter and CMB power spectra for
accelerated Bayesian inference.

The physics of CMB angular power spectra is well understood and physics-based approximations for the various  physical effects
have been developed~\citep{hu1994anisotropies}.  These enabled the first
 interpolation-based approaches such as DASh~\citep{Kaplinghat:2002mh}, and PICO \citep{Fendt_2007_Pico1,Fendt_2007_Pico2} which provided rapid evaluation of likelihoods for early Planck analyses. 
The first machine learning approaches emerged from bottleneck speed-ups in EB codes such as CosmicNet I \citep{albers2019cosmicnet} and later on II for extended cosmologies \citep{gunther2022cosmicnet}.

(Auto-)differentiable emulators such as \texttt{capse.jl} \citep{Bonici2024Capse} offer substantial improvements for inference, since one can leverage gradient-based methods such as Pathfinder or Hamiltonian Monte Carlo to speed up parameter optimisation or sampling. 
This is enabled by the development of differentiable likelihoods \citep{balkenhol2024candl} that do not rely on finite-difference methods and guarantee stable Fisher matrices.

Extending the cosmological model beyond $\Lambda$CDM, \citet{bolliet2024high} introduced a set of neural network emulators based on the CosmoPower framework \citep{spurio2022cosmopower}, which individually make predictions for pure $\Lambda$CDM, wCDM, and $\Lambda$CDM with massive neutrinos.
There is, however, currently no single unified CMB lensing-potential emulator that encapsulates the full joint parameter space of $w_0 w_a$ and massive neutrinos. 
This emulator suite has been used in the recent ACT DR6 analysis \citep{AtacamaCosmologyTelescope:2025nti},
demonstrating that emulators are indeed being used for inference, although currently in conjunction with precise predictions from CAMB and CLASS.

In recent years, symbolic regression (SR), the method of finding an analytical function to fit a dataset rather than relying on one of the standard machine learning ``black-box'' models, has risen in popularity. 
For example, the \textsc{syren} suite of emulators encapsulates high-precision predictions of the matter power spectrum, whether that be linear \citep{bartlett2023precise} or nonlinear \citep{bartlett2024syren} for $\Lambda$CDM. 
Similar approaches were used to extend these to cosmologies which include dynamical dark energy and massive neutrinos \citep{Sui:2024wob}, and most recently to incorporate the contribution due to baryon feedback  \citep{Kammerer:2025dbi}. 
Emulation of the linear matter power spectrum has also been extended to the Generalized Dark Matter model\citep{vokrouhlicky2026}. 
 For the mildly non-linear regime, SR has been used to significantly improve the computational speed of corrections to the matter power spectrum in the Effective Field Theory of LSS (EFTofLSS)
~\citep{Farakou:2025tuq}. The NN based approach of~\citep{Bonici:2025ltp} which also computes some of the EFTofLSS corrections, employed SR techniques for the computation of the growth factor.
The latter is emulated with SR with striking precision in~\cite{Bartlett_2025} which also shows how SR-based emulators can accelerate end-to-end inference in weak-lensing analyses.
Apart from cosmology, SR has  been used to emulate observables for beyond-standard-model phenomena
that are difficult to calculate \citep{abdussalam2025royalsociety,abdussalam2025symbolic} and collider physics \citep{bahl2025mathcal}. 

In this paper, we employ  SR  to create an easy-to-use and accurate suite of CMB emulators. We create emulators for a widely used extension of the $\Lambda$CDM model which includes 
 dynamical dark energy and massive neutrinos.
Dynamical dark energy has generated attention over the last year or so with the release of the DESI Y2 data~\citep{adame2025desi}.
The Chevallier-Polarski-Linder (CPL) parametrization appears sufficient to capture the improvement in fit, 
with the reduction in $\chi^2$ saturating once two dark-energy parameters are introduced in the analyses \citep{lodha2025extended}. We thus include the two CPL parameters $w_0$ and $w_a$ into our emulator.
 Meanwhile, CMB lensing is affected by the mass of neutrinos, and therefore acts as a complementary probe to oscillation experiments.
The fact that tensions between datasets, such as Planck, and LSS surveys such as DES \citep{adame2025desi,DES:2022oqz} or KiDS \citep{KiDS:2020suj}, 
can result in statistical artifacts in the form of a preference for unphysical negative mass values \citep{green2025cosmological} shows the sensitivity of current data to neutrino mass. 
We thus, include the sum of neutrino masses (henceforth denoted with $\Sigma m_\nu$) as a further parameter.

\begin{figure}
\centering
    \includegraphics[width=\linewidth]{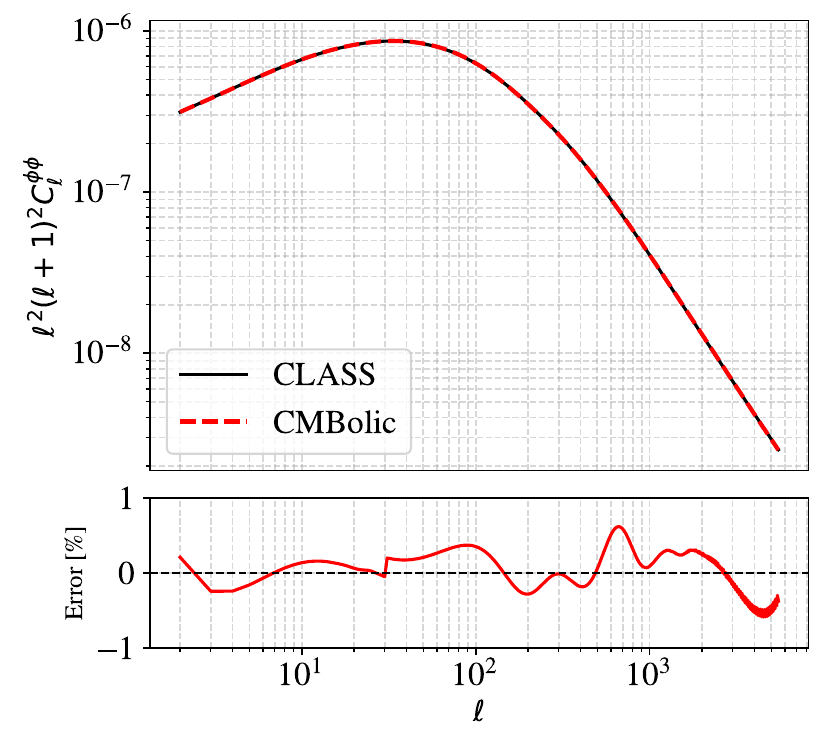}
    \caption{Comparison between CLASS and CMBolic for a Planck 2018 best-fit \(\Lambda\)CDM cosmology, with \(w_0=-1\), \(w_a=0\), and \(m_\nu=0.06\,{\rm eV}\). The upper panel shows the scaled lensing potential spectrum and the lower panel shows the fractional emulator error, which is well below the percent-level accuracy target motivated by lensing-noise forecasts.}
    \label{fig:error_fiducial}
\end{figure}

We note that, to the best of our knowledge, CMBolic is the only package
which covers the joint $\{w_0, w_a, \Sigma m_\nu\}$ parameter space with a single emulator. A simultaneous treatment of these parameters is necessary to account for potential degeneracies in data from future experiments. Alongside this relevance to current cosmology, there are many other benefits of the symbolic approach, such as seamless inclusion into any code, irrespective of programming language or requirements on external packages.
As a first illustration of the target emulated by CMBolic, Fig.~\ref{fig:error_fiducial} compares the emulator prediction with CLASS for a Planck 2018 best-fit \(\Lambda\)CDM cosmology, with the extended parameters fixed to \(w_0=-1\), \(w_a=0\), and \(m_\nu=0.06\,{\rm eV}\). The lower panel shows the fractional residual of the symbolic expression relative to the Boltzmann-code prediction. The residuals are well below the percent-level accuracy target motivated below by current and Stage-4-like lensing-noise forecasts.

This paper is structured as follows: in Sects.~\ref{sec:lensing} and \ref{sec:Symbolic} we briefly introduce CMB lensing and symbolic regression, in Sect.~\ref{sec:emulator} we specify training techniques and dataset generation with all due considerations. Sect.~\ref{sec:performance} contains analyses of the emulator performance and a comparison with other available CMB emulators, and we conclude in Sect.~\ref{sec:conclusions}. 
We assume a spatially flat background and units where the speed of light is unity.


\section{CMB Lensing}\label{sec:lensing}

CMB lensing is the deflection of CMB photon trajectories by the intervening large-scale structure (LSS). 
The observed lensed CMB field  
$\tilde{X}(\hat{n})$ is related to the unlensed field $X(\hat{n})$ by a remapping: $\tilde{X}(\hat{n})=X(\hat{n}+\nabla\phi(\hat{n}))$, where $\phi(\hat{n})$ is the CMB lensing potential, representing the line-of-sight integral of the Weyl gravitational potential $\Psi$:
\begin{align}
    \phi(\hat{n})=-2 \int_0^{\chi_*} {\rm d} \chi \, \frac{\chi_*-\chi}{\chi_* \chi} \Psi\left(\chi \hat{n}, \eta_0-\chi\right).
\end{align}
Above,  $\chi$ is the comoving distance, $\chi_*$ the comoving distance to the last-scattering surface and $\eta_0$  the conformal time today.
The potential $\phi(\hat{n})$ is characterised by its angular power spectrum
\begin{equation}
    \langle\phi_{\ell m} \phi_{\ell' m'}^*\rangle = \delta_{\ell\ell'} \delta_{mm'} C_\ell^{\phi\phi}, \label{eq:clphiphi_def}
\end{equation}
which defines the $ C_\ell^{\phi\phi}$ coefficients.
The full expression involves integration over wavenumber $k$ and a double integral over the line of sight:
\begin{equation}
    \label{eq:clphiphi_full}
    \begin{split}
        C_\ell^{\phi\phi} = 16\pi & \int \frac{ {\rm d}k}{k} \int_0^{\chi_*}  {\rm d}\chi \int_0^{\chi_*}  {\rm d}\chi' \mathcal{P}_{\Psi}\!\left(k;\eta_0-\chi,\eta_0-\chi'\right) \\
    & \times  j_\ell(k\chi) j_\ell(k\chi') \left(\frac{\chi_* - \chi}{\chi_* \chi}\right) \left(\frac{\chi_* - \chi'}{\chi_* \chi'}\right),
    \end{split}
\end{equation}
where $j_\ell$ are the spherical Bessel functions and $\mathcal{P}_{\Psi}(k;\eta,\eta')$ is the dimensionless unequal-time power spectrum of the Weyl potential. Under the Limber approximation \citep{limber1953analysis}, which is accurate for large multipoles ($\ell \gg 1$), the power spectrum is given by a single integral
\begin{equation}
    C_\ell^{\phi\phi} \approx \frac{8\pi^2}{\nu^3} \int_0^{\chi_*}  {\rm d}\chi \, \chi \left(\frac{\chi_* - \chi}{\chi_* \chi}\right)^2 \mathcal{P}_{\Psi}\left(k=\frac{\nu}{\chi};\eta_0-\chi\right),
\;
\nu=\ell+\frac12 .
\label{eq:cl_limber}
\end{equation}
Typically the switch to the Limber approximation happens around $\ell=30$.
For practical computation, it is often more convenient to work with the total matter power spectrum, $P(k)$, which in a flat universe is related to
the dimensionless Weyl-potential power spectrum through
\begin{equation}
    \mathcal{P}_{\Psi}(k; \eta) = \frac{9 \Omega_{\rm m}^2(\eta) \mathcal{H}^4(\eta)}{8\pi^2 k} P(k; \eta), \label{eq:ppsips}
\end{equation}
where $\Omega_{\rm m}(\eta)$ is the matter density fraction and $\mathcal{H}(\eta)$ is the conformal Hubble parameter as functions of conformal time $\eta$.
Although CMB lensing constrains all cosmological parameters to some degree, it is most sensitive to 
the parameters that govern the late-time growth of structure, such as the amplitude of matter fluctuations, $\sigma_8$, and the total matter density, $\Omega_{\rm m}$.


\section{Symbolic regression}\label{sec:Symbolic}
Symbolic regression is a technique in machine learning focused on uncovering explicit mathematical expressions that model the underlying relations in a dataset \citep[see, e.g.][for a recent review]{Kronberger_2024}. In contrast to traditional regression methods, which require the specification of a fixed functional form in advance, symbolic regression searches the space of possible mathematical expressions to discover both the structure and parameters of the model simultaneously. This avoids imposing a single predetermined functional form, although the resulting expressions still depend on choices such as the operator set, complexity measure, training distribution, and stochastic search procedure. The main advantages of the symbolic approach are that the final models are explicit, portable, differentiable, straightforward to reproduce, and easy to deploy as closed-form functions without relying on specialized machine-learning frameworks.

Due to SR being NP-hard \citep{virgolin2022symbolic}, an exhaustive search for the optimal expression is computationally infeasible if that expression is sufficiently complex. Furthermore, the discrete and non-differentiable nature of the formula search space rules out gradient-based optimization. The problem thus requires a heuristic method to find high-quality, ``good-enough'' solutions in a practical amount of time, with genetic programming (GP) \citep{10.5555/138936} being the state-of-the-art approach. In this approach, equations represented by trees are evolved in analogy to natural selection towards populations with the best fit. Usually, SR is a multi-objective optimization problem, as we want to encode the information accurately in the simplest possible formula. The result of this genetic procedure is therefore a Pareto front of the best-fit models for each complexity. 

Several successful implementations of SR are available today with various regions of applicability.
Due to the explosion of applications of SR in the physical sciences, advances in the development of tools and the theoretical background have also accelerated. 
Recent advances include combining SR with large language model approaches \citep{grayeli2024symbolic}, GPU optimized approaches \citep{muthyala2025symantic}, and significant improvements in efficiency of parameter space search using equality saturation \citep{de2025improving}. These advances establish SR as a compelling tool for complex problems in physics, including the emulation of CMB lensing.

In this work, we have chosen to use the GP-based code Operon \citep{10.1145/3377929.3398099}. Operon has consistently demonstrated strong performance and efficiency across a variety of symbolic regression tasks \citep{DBLP:journals/corr/abs-2107-14351, radwan2024comparison}, and we have confirmed this with our own testing. We selected Operon for this work due to its efficiency, robustness and its scaling with larger training datasets.
We tried various available symbolic regression suites such as PySR \citep{cranmer2023interpretable}, which worked well only on much smaller datasets than are needed here, but found that only Operon had the necessary efficiency to achieve a reasonable fit. Curiously, in a cosmology-specific setting, Operon was reported to underperform other algorithms \citep{Thing_2025}. This is, however, due to the very small chosen datasets of order 10-1000, where speed is not the bottleneck of the GP procedure.

Operon's unique features include its computational efficiency due to a linear tree representation and modern C++ implementation. It uses the Balanced Tree Creator (BTC) for diverse population initialization and supports fitness evaluation using dual numbers for automatic differentiation. This allows integration with gradient-based local search methods for parameter optimization. The code is also designed for parallel execution.
Operon uses length as a measure of complexity, which is calculated as the total number of nodes in the expression's tree representation, which includes all operators, variables, and constants.

\section{Lensing emulator}\label{sec:emulator}
\subsection{Dataset generation}
Our symbolic emulator is trained on high-accuracy spectra computed with the CLASS \citep{lesgourgues2011cosmic} code version v3.3.0 with ACT DR6-inspired accuracy settings listed in Appendix~\ref{sec:appendix_A} \citep{calabrese2025atacama}. We create a Latin hypercube (LH) of sample cosmological parameters drawn from uniform priors in the ranges listed in Table \ref{table:parameter_ranges}. These parameter ranges define a pragmatic training domain for lensing-only inference in extended cosmologies. They are motivated by the broad domains used in recent symbolic matter-power emulators for \(w_0w_a+\Sigma m_\nu\) cosmologies, while being restricted enough for a single compact symbolic expression to reach sub-percent accuracy. The central
region of this hypercube contains current CMB and BAO posterior constraints, and the full volume covers the priors used in our ACT/Planck lensing-only inference tests below.

\begin{table}
\caption{Parameter ranges for the LH training and validation data}
\label{table:parameter_ranges}
\centering
\begin{tabular}{lcc}
\hline\hline
Parameter & Lower bound & Upper bound \\
\hline
$10^9 A_{\rm s}$ & 1.7  & 2.5  \\
$\omega_{\rm m}$ & 0.11 & 0.18 \\
$\omega_{\rm b}$ & 0.02 & 0.024 \\
$h$              & 0.63  & 0.77  \\
$n_{\rm s}$      & 0.92 & 1.01 \\
$w_0$            & -1.3 & -0.7 \\
$w_a$            & -0.7 & 0.5 \\
$m_\nu$ [eV]     & 0    & 0.3 \\
\hline
\end{tabular}
\end{table}

Even though some existing emulators target the higher multipole ranges of proposed experiments such as CMB-HD \citep{sehgal2019cmb} which were planned to increase the very high $\ell$ sensitivity to $10^4$, we limit ourselves to upcoming experiments as the others are now unfortunately postponed. We emulate the spectrum over \(2\leq \ell \leq 5500\), which covers the range
relevant for the current lensing likelihoods and Stage-4-like forecast
comparisons considered in this work. We generated a training set of 150 cosmologies and a  same-size validation set, which were sufficient to obtain the desired precision.

The optical depth to reionization $\tau$ is fixed at 0.055 for all cosmologies in our sample. It affects the primary CMB anisotropies directly, but has negligible impact on \(C_\ell^{\phi\phi}\) at fixed primordial amplitude and late-time cosmological parameters. We tested this assumption by varying \(\tau\) over \(0.03\leq\tau\leq0.08\)
at fixed \(A_{\rm s}\) and otherwise fixed cosmological parameters. The
resulting change in \(C_\ell^{\phi\phi}\) was below \(0.16\%\) over
\(2\leq\ell\leq5500\), smaller than the emulator validation error. We also tested this assumption by varying this parameter in the same range in a separate training set and running symbolic regression on this set. The expressions did not contain $\tau$ as a parameter until one considered expressions of very high complexity, where the dependence was through negligible contributions due to bloat effects of genetic algorithms.  

We consider only one species of massive neutrinos, as is assumed in the latest ACT analysis \citep{calabrese2025atacama}. The sampled parameter \(m_\nu\) is therefore numerically equal to the total neutrino mass \(\Sigma m_\nu\). Many of the aforementioned emulators use the three degenerate species approximation, as it has been shown to be indistinguishable from specific hierarchies even with Stage 4 data \citep{archidiacono2020will}. 

The recommended accuracy settings for CLASS from the ACT DR6 analysis are listed in Appendix \ref{sec:appendix_A}.
Only versions of CLASS 3.3 or later have sufficient lensing accuracy to match CAMB \citep{lewis2011camb} contours on ACT DR6 lensing due to an updated extrapolation of the lensing integral and updates to the latest version of HMcode \citep{mead2021hmcode}.

We have tested fitting various commonly used versions of lensing spectra such as pure $C_\ell$, $\ell(\ell+1)C_\ell$, or $\ell^2(\ell+1)^2C_\ell$. The second variant yielded the best results as it is a monotonic function of $\ell$ with the simplest dependence. We have tested using either $\ell$ or $\log \ell$ as independent parameters, with $\log \ell$ yielding a superior fit for relevant scales.
Due to the poor scaling of symbolic regression with the number of training points, we did not train on the full CLASS output grid. Instead, for each cosmology we first computed the spectrum on 5000 multipoles spanning $\ell=2$ to $\ell=5500$, and then selected 500 points uniformly spaced in $\log \ell$ over this range. By experimentation, this logarithmically spaced 500-point subset provided an optimal compromise between accuracy and training time. The total number of training points was therefore 75\,000. The validation set, by contrast, was evaluated on the full 5000-point multipole grid.

Inspired by similar symbolic emulation work, we attempted to factorize $C_\ell$ into a product of the $\Lambda$CDM contributions and a correction for $w_0w_am_\nu$ cosmologies. However,
 the accumulated error extended beyond our target precision. Hence, we chose to only factor out $A_{\rm s}$ as expected from linear theory
\begin{align}
    C_\ell=A_{\rm s} \times F_{\ell}(A_{\rm s},n_{\rm s},h, \omega_{\rm m}, \omega_{\rm b}, w_0, w_a, m_\nu).
\end{align}
We tested runs without any other dependence on $A_{\rm s}$ for parameter efficiency. The fit is, however, improved with the addition of $A_{\rm s}$ in other ways than just a linear prefactor. This dependency arises due to effects of using the nonlinear power spectrum prescription.  

We ran Operon for 24 hours on a 128-core machine to obtain our best Pareto fronts. This corresponds to roughly $10^5$ generations of evolution. The set of possible functions to use as basis operators was power, exponential, logarithm, square-root, analytical quotient (${\rm aq}(x,y) \equiv x / \sqrt{1 + y^2}$), as well as standard arithmetic operators ($+$, $-$, $\times$, $\div$). We also added $\cos$ as an operator for better precision.
To avoid spurious oscillatory fits, we rejected expressions containing nested trigonometric functions or trigonometric factors with periods that would be much shorter than the width of the parameter intervals. In terms of regression settings that were changed from default values we set the maximum expression
length to 120, coefficient optimizer steps to 3, and modified the  convergence threshold to \(\epsilon=10^{-3}\). The maximum generation and evaluation limits were set high enough that the wall-time limit was the effective stopping criterion.
 
As objectives, we used complexity and the root mean square error (RMSE).  This has proven to be marginally better than other metrics such as mean absolute error (MAE) or mean squared error (MSE). Due to the stochastic nature of SR, the variance of the performance of subsequent runs with the same hyperparameters is greater than changes of the parameters themselves around a reasonable local optimum. We therefore ran Operon 5 times and chose the best Pareto front, which is presented in Sect.~\ref{sec:pareto}.

\subsection{Nonlinear power spectrum}
The lensing potential power spectrum $C_\ell^{\phi\phi}$ is sensitive to the matter power spectrum, requiring the use of a nonlinear prescription, especially for small scale anisotropies. 

\begin{figure}
    \centering
    \includegraphics[width=\linewidth]{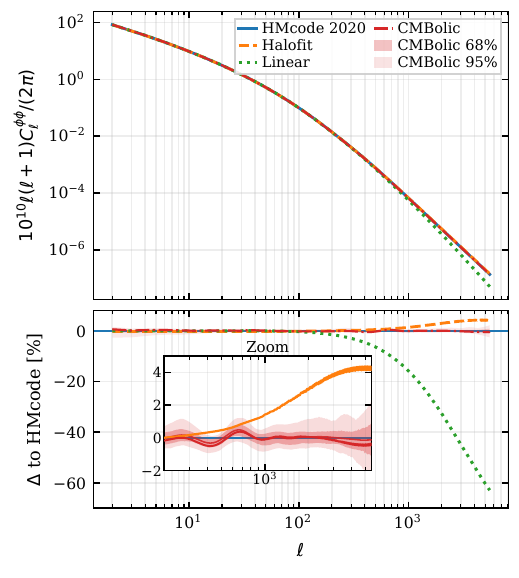}
    \caption{Comparison of nonlinear matter-power prescriptions for a Planck-like fiducial cosmology. The upper panel shows the scaled lensing-potential spectrum for HMcode 2020, Halofit, the linear matter power spectrum, and CMBolic. The lower panel shows fractional differences relative to HMcode 2020, the prescription used to generate the CMBolic training data. The CMBolic residual for the fiducial cosmology and the 68\% and 95\% validation residual bands, shown in red and highlighted in the inset, are much smaller than these modelling differences.}
    \label{fig:nonlinear}
\end{figure}

For a Planck-like fiducial cosmology, replacing HMcode 2020 by the linear matter power spectrum while keeping all other CLASS settings fixed changes \(C_\ell^{\phi\phi}\) by up to \(63\%\) over \(2\leq\ell\leq5500\), 
measured relative to the HMcode 2020 prediction, with the maximum occurring at \(\ell=5500\). The effect increasingly grows at high multipoles: 
the linear prediction is lower than HMcode 2020 by \(15.8\%\), \(32.5\%\), and \(60.3\%\) at \(\ell=1000\), \(2000\), and \(5000\), respectively. This is shown in Fig. \ref{fig:nonlinear} alongside the error contours of CMBolic. These can be seen as much lower than the differences in the nonlinear modelling prescriptions.

Standard nonlinear prescriptions include Halofit \citep{smith2003stable, takahashi2012revising} and HMcode \citep{mead2015accurate, mead2016accurate, mead2021hmcode}. More recently, symbolic emulators like \textsc{syren} \citep{bartlett2024syren, Farakou:2025tuq},  have also become available. While the differences between these modern nonlinear prescriptions (HMcode 2020, \textsc{syren}) are relatively small, reaching up to approximately $5\%$ for $C_\ell^{\phi\phi}$ within our $\ell$ range, the choice can still be relevant. We adopted HMcode 2020 \citep{mead2021hmcode} as implemented in CLASS v3.3 for generating our training data, primarily because it was the prescription used in the ACT DR6 analysis \citep{calabrese2025atacama} and offers validity across a slightly wider parameter space than currently available symbolic alternatives like \textsc{syren}, particularly regarding neutrino mass priors. In future work, it would be interesting to test whether symbolic nonlinear prescriptions such as \textsc{syren} can be incorporated directly into the pipeline, enabling an even more fully symbolic treatment of the lensing calculation.

\subsection{Target accuracy for Stage-4-like forecasts}

\begin{figure}
    \centering
    \includegraphics[width=\linewidth]{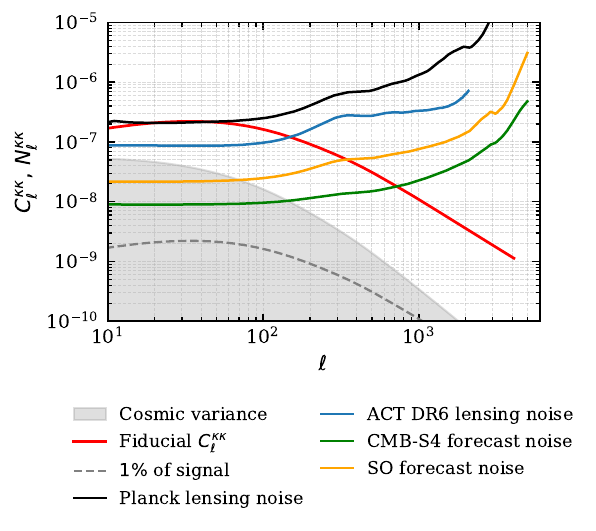}
    \caption{Lensing-reconstruction noise curves in the convergence convention, \(C_\ell^{\kappa\kappa}=\ell^2(\ell+1)^2C_\ell^{\phi\phi}/4\), compared with a fiducial lensing spectrum and a \(1\%\) fractional-error reference curve. The Planck and ACT DR6 curves correspond to existing lensing reconstructions, while the SO and CMB-S4 curves are forecast curves. Since the convergence and lensing-potential conventions differ only by a common multipole-dependent rescaling, the fractional comparison to the emulator error is unchanged.}
    \label{fig:noise_norm_t}
\end{figure}

In Fig.~\ref{fig:noise_norm_t}, we plot representative lensing-reconstruction noise curves for Planck, ACT DR6, SO, and CMB-S4, together with a fiducial convergence spectrum and a \(1\%\) fractional-error reference curve.
The Planck curve is based on the Planck 2018 lensing reconstruction \citep{aghanim2020planck}, while the ACT DR6 curve is based on the ACT DR6 lensing likelihood and associated data products \citep{ACT:2023dou,ACT:2023kun}. For the ACT+Planck combination used below we also cite \citet{carron2022cmb}. For SO we use the public \texttt{so\_noise\_models} lensing-noise release, originally associated with the Simons Observatory forecasts \citep{ade2019simons}, and for the Stage-4-like curve we use a forecast based on the CMB-S4 reference design \citep{abazajian2019cmb}.\footnote{The SO lensing-noise file used here is \url{https://github.com/simonsobs/so_noise_models/blob/master/LAT_lensing_noise/lensing_v3_1_1/nlkk_v3_1_0_deproj1_SENS2_fsky0p4_it_lT30-3000_lP30-5000.dat}.}
We use the \(1\%\) curve as a baseline validation target rather than as a strict requirement on the mean fractional error alone. In addition to the mean absolute fractional error, we inspect the \(\ell\)-dependent 68\% and 95\% validation bands and require them to remain approximately within the percent-level range over the multipoles most relevant for the likelihood. This prevents a model
with a small average error from introducing localized residuals at specific multipoles to which a lensing likelihood could be sensitive. In practice the needed RMSE is then smaller than the 1\% target.

Low multipoles are dominated by cosmic variance, while high multipoles are dominated by instrumental and reconstruction noise. The \(1\%\) reference curve lies below these representative statistical error scales over the range relevant for the lensing applications considered here. This motivates our adoption of a percent-level target accuracy, and in practice we use a 1\% baseline criterion when selecting the final symbolic model.

\subsection{Formula post-processing}

After choosing a function from Operon, we simplified it by removing unnecessary terms such as offsets that are of lower magnitude than our target accuracy, terms that do not meaningfully contribute, and protected division in cases where it has negligible effect. We also rounded two coefficients that were less than $10^{-5}$ away from unity and therefore remove them from the formula. We then refitted the coefficients of our simplified formula on the training dataset. This procedure reduced the complexity of our emulator by 6 without sacrificing accuracy. Lastly, we rounded all coefficients to 4 significant figures and repeated all  tests. 
We attribute the failure of finding this expression automatically to traditional GP concepts such as bloat. 
This process of employing physically informed simplification represents the strength of SR over other non-symbolic ML approaches such as NNs. We did not explicitly impose known low- or high-$\ell$ asymptotic limits during the simplification step; instead, the behavior of the final expression was checked a posteriori against the CLASS output.

\subsection{Pareto fronts and convergence}\label{sec:pareto}
\begin{figure}
    \centering
       \includegraphics[width=\linewidth]{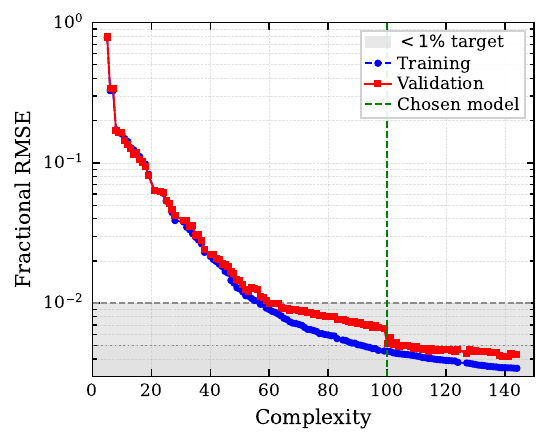}
    \caption{Pareto fronts of the leading Operon run for lensing anisotropies on training and validation data. The shaded region roughly indicates a 1\% criterion for accuracy; the vertical line shows the chosen model.}
    \label{fig:pareto}
\end{figure}

To examine the convergence and overfitting, we plot the Pareto front in Fig.~\ref{fig:pareto} -- the best fit expression based on complexity -- on both the training and validation dataset. We see that the training and validation losses do not diverge at our chosen model. In this case, overfitting in between discrete values of $\ell$ is not an issue as would be in the case of a continuous variable. To test this explicitly, we evaluated the final emulator on the full validation grid containing all 5000 multipoles, not only on the logarithmically subsampled multipoles used during training; the resulting error statistics are discussed in Sect.~\ref{sec:emulator_precision}.

Our chosen model is indicated in Fig.~\ref{fig:pareto} by the vertical line, and we plot
a horizontal $1\%$ RMSE baseline model quality criterion. For practical use, we want at least the 2-sigma band to be roughly confined under $1\%$, 
which for a Gaussian error distribution would necessitate the RMSE to be $0.5\%$. We therefore chose a model where the validation RMSE approaches this value and does not significantly diverge from the training RMSE. 
We can see that getting the spectrum to roughly $10\%$ precision can be done with a function of complexity under 20,
but to go an order of magnitude further towards a useful emulator already requires  much higher complexity functions.

\subsection{Full formula}\label{sec:formula}
Our emulator formula for $C_\ell^{\phi\phi}$ is
\begin{align}
&
\log\left(10\ell(\ell+1)\frac{C_\ell^{\phi\phi}}{2\pi A_{\rm s}}\right)  =
  \frac{C_{0} n_{\rm s} + \frac{C_{1} x}{\sqrt{C_{2} x^{2} + 1}}}{\sqrt{\cos^{2}{\left(C_{3} x \right)} + 1}} 
+\frac{ \left(C_{4} x + C_{5}\right)}{\sqrt{C_{6} n_{\rm s}^{2} + 1}} 
\nonumber
\\
&
\ \  \ \
- \frac{ \left(C_{7} \omega_{\rm m} + \frac{C_{8}}{\sqrt{C_{9} h^{2} + 1}} + C_{10} m_{\nu} + C_{11} x + C_{12} \omega_{\rm b}\right) e^{C_{13} \tilde A_{\rm s} + C_{14} x}}{\sqrt{\cos^{2}{\left(C_{15} \tilde A_{\rm s}  + C_{16} x \right)} + 1}}
\nonumber
\\
&
\ \ \ \
 - \frac{\left( C_{17} x + C_{18} w_a + C_{19} w_0\right) \cos{\left(C_{20} x \right)} - e^{C_{21} w_a} - e^{C_{22} w_0 + C_{23} w_a}}{\sqrt{\left(C_{24} \omega_{\rm m} + C_{25} w_0 + C_{26} h + C_{27} w_a\right)^{2} + 1}} 
\nonumber
\\
&
\ \  \ \
+ \frac{\left(C_{28} \omega_{\rm b} + C_{29} x - \cos{\left(C_{30} + C_{31} x \right)}\right)}{\sqrt{\left(C_{32} \omega_{\rm b} + C_{33} m_{\nu} + C_{34} w_0 + C_{35} \omega_{\rm m}\right)^{2} + 1}} 
\nonumber
\\
&
\ \  \ \
+ \frac{\frac{C_{36}}{\sqrt{\left(C_{37} x + C_{38} m_{\nu}\right)^{2} + 1}} + C_{39} m_{\nu} + \left(C_{40} n_{\rm s} + C_{41} \omega_{\rm m}\right)^{C_{42} \omega_{\rm b}}}{\sqrt{\left(C_{43} + C_{44} x\right)^{2} + 1}},
\label{equation_90}
\end{align}
where $x\equiv\log\ell$ and $\tilde A_{\rm s}=10^9A_{\rm s}$.  The coefficients $C_i$  are 

\begin{align*}
C_{0} &= -1.702 & C_{1} &= 1.057 & C_{2} &= 0.1021 \\
C_{3} &= 0.3568 & C_{4} &= 2.258 & C_{5} &= -7.782 \\
C_{6} &= 0.7468 & C_{7} &= -1.107 & C_{8} &= -32.47 \\
C_{9} &= 9.496\times10^4 & C_{10} &= 0.09044 & C_{11} &= 0.1950 \\
C_{12} &= 4.975 & C_{13} &= -0.01905 & C_{14} &= 0.1686 \\
C_{15} &= 0.06812 & C_{16} &= 0.2026 & C_{17} &= -0.03212 \\
C_{18} &= 0.1698 & C_{19} &= 0.2092 & C_{20} &= 0.6773 \\
C_{21} &= -0.2246 & C_{22} &= 7.950 & C_{23} &= 8.180 \\
C_{24} &= 11.64 & C_{25} &= -4.975 & C_{26} &= -4.554 \\
C_{27} &= -2.642 & C_{28} &= -17.43 & C_{29} &= 0.5517 \\
C_{30} &= -209.6 & C_{31} &= 0.5644 & C_{32} &= -33.92 \\
C_{33} &= -0.5488 & C_{34} &= -0.05430 & C_{35} &= 18.37 \\
C_{36} &= -1.286 & C_{37} &= 0.1703 & C_{38} &= -1.840 
\end{align*}
\begin{align*}
C_{39} &= 0.8291 & C_{40} &= -0.4749 & C_{41} &= 5.194 \\
C_{42} &= 3.640 & C_{43} &= 11.42 & C_{44} &= -2.083,
\end{align*}

\noindent and were determined by the fitting procedure described above.
Here and throughout, \(\log\) denotes the natural logarithm and \(m_\nu\) is expressed in eV. The left-hand side of Eq.~\eqref{equation_90} is normalized by the physical primordial amplitude \(A_{\rm s}\), while the right-hand side uses the rescaled variable \(\tilde A_{\rm s}\equiv 10^9A_{\rm s}\).

Eq.~\eqref{equation_90} should be interpreted as an emulator rather than as a physically transparent decomposition of the lensing spectrum. Its analytic form is nevertheless useful: it is compact, differentiable, and involves only standard mathematical operations.  The trigonometric factors do not introduce rapid oscillations over the trained range.
The fitted interval has width \(\Delta x=\log(5500/2)=7.92\). The shortest retained trigonometric period is \(\pi/C_{3}=8.80\), from the
\(\cos^2(C_{3}x)\) term; the remaining periods are \(2\pi/C_{20}=9.28\), \(2\pi/C_{31}=11.13\), and \(\pi/C_{16}=15.51\). Thus the trigonometric factors span only \(0.51\)--\(0.90\) cycles over the full training interval in \(x=\log\ell\). They therefore act as broad smooth curvature corrections rather than as rapidly oscillating features.
Due to the nature of analytical quotients, i.e. denominators in the form of square roots appear as the sum of unity and a square number, no poles can appear. The only non-integer power,
\[
\left(C_{40}n_{\rm s}+C_{41}\omega_{\rm m}\right)^{C_{42}\omega_{\rm b}},
\]
is also well defined throughout the training domain because the base remains positive for all sampled cosmologies.

The dependence on \(A_{\rm s}\) is dominated by the explicit normalization in Eq.~\eqref{equation_90}, with residual dependence on \(\tilde A_{\rm s}\) entering through the nonlinear matter-power prescription used to generate the training data. The dark-energy parameters enter through a small number of combinations, in particular through \(C_{22}w_0+C_{23}w_a\) and through linear combinations involving \(w_0\), \(w_a\), \(h\), and \(\omega_{\rm m}\), reflecting the broad degeneracy directions expected from lensing-only data. The neutrino-mass dependence is similarly smooth, entering through linear terms and analytic-quotient denominators rather than through rapidly varying functions. We did not impose analytic low- or high-\(\ell\) asymptotic limits during symbolic regression or post-processing; instead, the expression was validated directly against CLASS over the full training domain. 

For physical intuition, the low-multipole lensing spectrum is sourced by large, mostly linear modes of the Weyl potential, with a shape determined by the primordial tilt, the transfer function, and the lensing projection kernel. If the Weyl-potential power is locally approximated as \(\mathcal P_\Psi(k)\propto k^{n_{\rm s}-1}\), the Limber scaling would suggest \(C_\ell^{\phi\phi}\propto \ell^{n_{\rm s}-4}\), or \(\ell(\ell+1)C_\ell^{\phi\phi}\propto \ell^{n_{\rm s}-2}\). This argument is only heuristic at the lowest multipoles, where the full-sky calculation is required, and no such scaling was imposed in the fit. 
For a Planck-like fiducial cosmology, expanding \(F(x)\equiv\log[10\ell(\ell+1)C_\ell^{\phi\phi}/2\pi A_{\rm s}]\) around
\(x_0=\log 10\) gives\[ F(x)\simeq1.5167-1.5171(x-x_0)-0.1376(x-x_0)^2
-0.0177(x-x_0)^3 .\] Thus, over \(2\leq\ell\leq30\), the emulator behaves roughly as a local power law \(\ell(\ell+1)C_\ell^{\phi\phi}\propto \ell^{-1.5}\) with smooth logarithmic curvature corrections. Over \(2\leq\ell\leq30\), the local
slope \(dF/d\log\ell\) varies smoothly from about \(-1.2\) to \(-1.9\).
At high multipoles, the spectrum is shaped by the matter transfer function and nonlinear growth, so no simple universal power-law limit is expected over the range relevant for CMB lensing.

The emulator should therefore only be used within the parameter ranges in Table~\ref{table:parameter_ranges} and the multipole interval \(2\leq\ell\leq5500\). In particular, Eq.~\eqref{equation_90} should not be extrapolated to arbitrarily large \(\ell\), where its symbolic high-\(\ell\) behavior need not reproduce the physical asymptotic form of the lensing spectrum.

\subsection{Emulator precision}\label{sec:emulator_precision}
\begin{figure*}
    \centering
    \includegraphics[width=\linewidth]{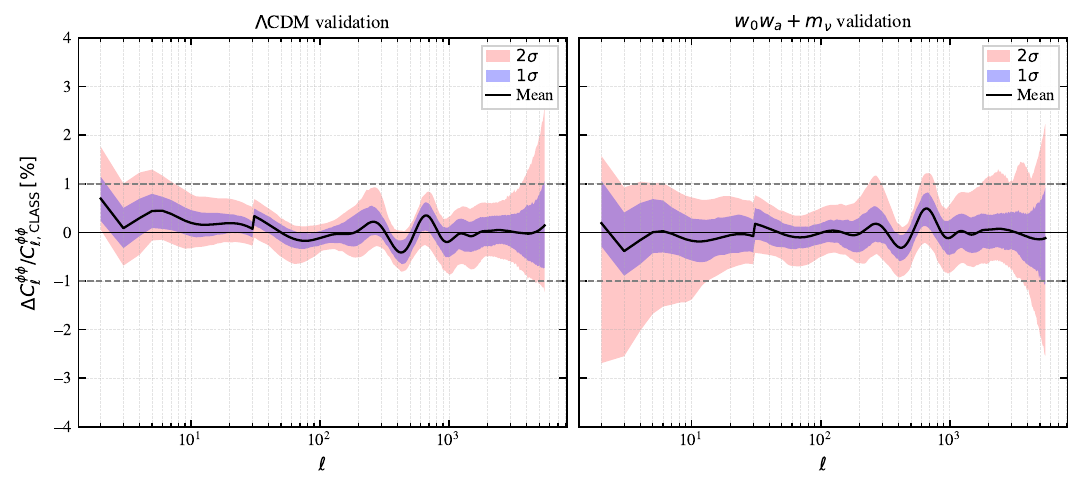}
    \caption{Fractional emulator-error bands evaluated on the validation datasets. The shaded regions show the central 68\% and 95\% intervals across validation cosmologies at fixed \(\ell\). Dashed horizontal lines mark the \(1\%\) baseline accuracy target motivated by Fig.~\ref{fig:noise_norm_t}. Left: pure \(\Lambda\)CDM cosmologies. Right: the full extended cosmology. 
The small discontinuity at  $\ell=30$ is due to switching on/off the Limber approximation when computing the spectrum with CLASS, in both the training and validation data.}
    \label{fig:emulator_precision}
\end{figure*}

We evaluate the performance of the emulator on a validation dataset generated as a LH with the same priors and CLASS settings as the training dataset, containing all 5000 $\ell$-values as opposed to a logarithmically spaced subset used in the training dataset. In Fig.~\ref{fig:emulator_precision}, we look at the error distribution of our emulator for each $\ell$ and plot the one and two-sigma errors on this dataset, as well as on a pure $\Lambda$CDM dataset.  We see that the emulator is roughly confined to the desired $1\%$ error range except for the high $\ell$ tail, where it grows to $2\%$. Nevertheless, overall mean absolute fractional errors are $0.27\%$ and $0.32\%$ on the $\Lambda$CDM validation and extended cosmology validation datasets, respectively. The small discontinuity at $\ell=30$ is due to the switch to the Limber approximation in the data that propagates into a discontinuity in the error plots, but is smaller than the mean error. 

A more practical error analysis can be obtained by normalizing to optimistic noise forecasts for Stage 4 inference. Even with the constrained size, forecasts for S4 still represent desired precision of the next generation of CMB instruments and therefore the minimum requirements of any analysis tools.

\begin{figure*}
    \centering
    \includegraphics[width=\linewidth]{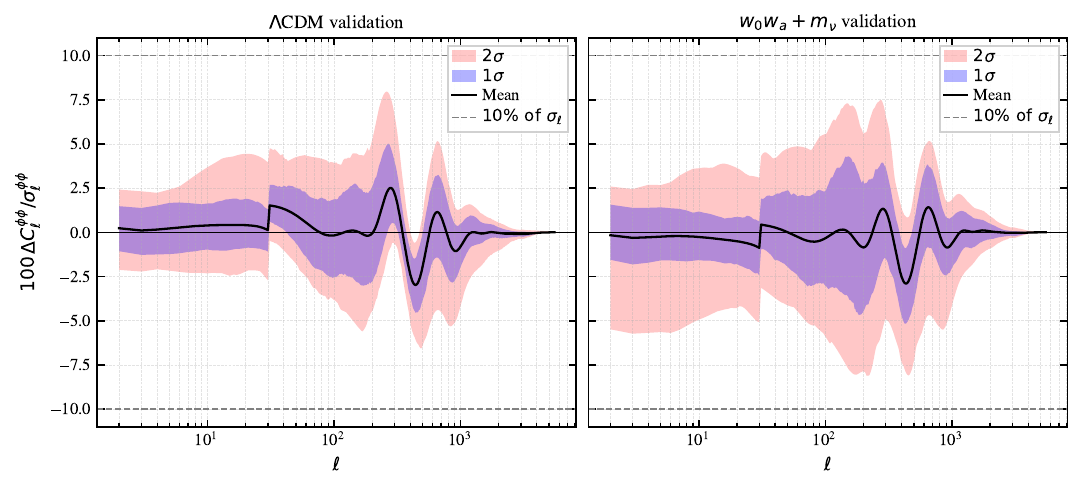}
    \caption{Validation-set emulator residuals normalized by the combined instrumental-noise and cosmic-variance uncertainty,
\(\Delta C_\ell^{\phi\phi}/\sigma_\ell^{\phi\phi}\) in percent, for the Stage-4-like forecast comparison. The shaded regions show the central 68\% and 95\% intervals across validation cosmologies. The emulator residuals remain well below \(10\%\) of the projected statistical uncertainty over the range relevant for the analysis. The small discontinuity at  $\ell=30$ is due to switching on/off the Limber approximation when computing the spectrum with CLASS, in both the training and validation data.}
    \label{fig:emulator_precision_sigma_normalized}
\end{figure*}

If we want a more holistic approach to errors, to not only include instrumental effects but also cosmic variance, we can use the Knox formula \citep{knox1995determination} 
\begin{align}
\sigma_{\ell}^{X X}=\sqrt{\frac{2}{f_{\mathrm{sky}}(2 \ell+1)}}\left(C_{\ell}^{X X}+N_{\ell}^{X X}\right),
\end{align}
where $f_{\rm sky}$ is the sky fraction covered and $N_\ell^{XX}$ is the instrument noise for each multipole and spectrum. 
Our emulator error divided by this noise is plotted in Fig.~\ref{fig:emulator_precision_sigma_normalized} in the idealized full-sky limit of $f_{\text{sky}}=1$. 
This represents the best case scenario for a survey which minimizes its noise, and therefore the worst-case scenario for our emulator.
Across the multipole range used in the likelihood comparison, the emulator residual remains well below \(0.1\,\sigma_\ell^{XX}\), even for the optimistic \(f_{\rm sky}=1\) case shown here, making CMBolic a valid emulator for next generation surveys.

\begin{figure*}
    \centering
    \includegraphics[width=\linewidth]{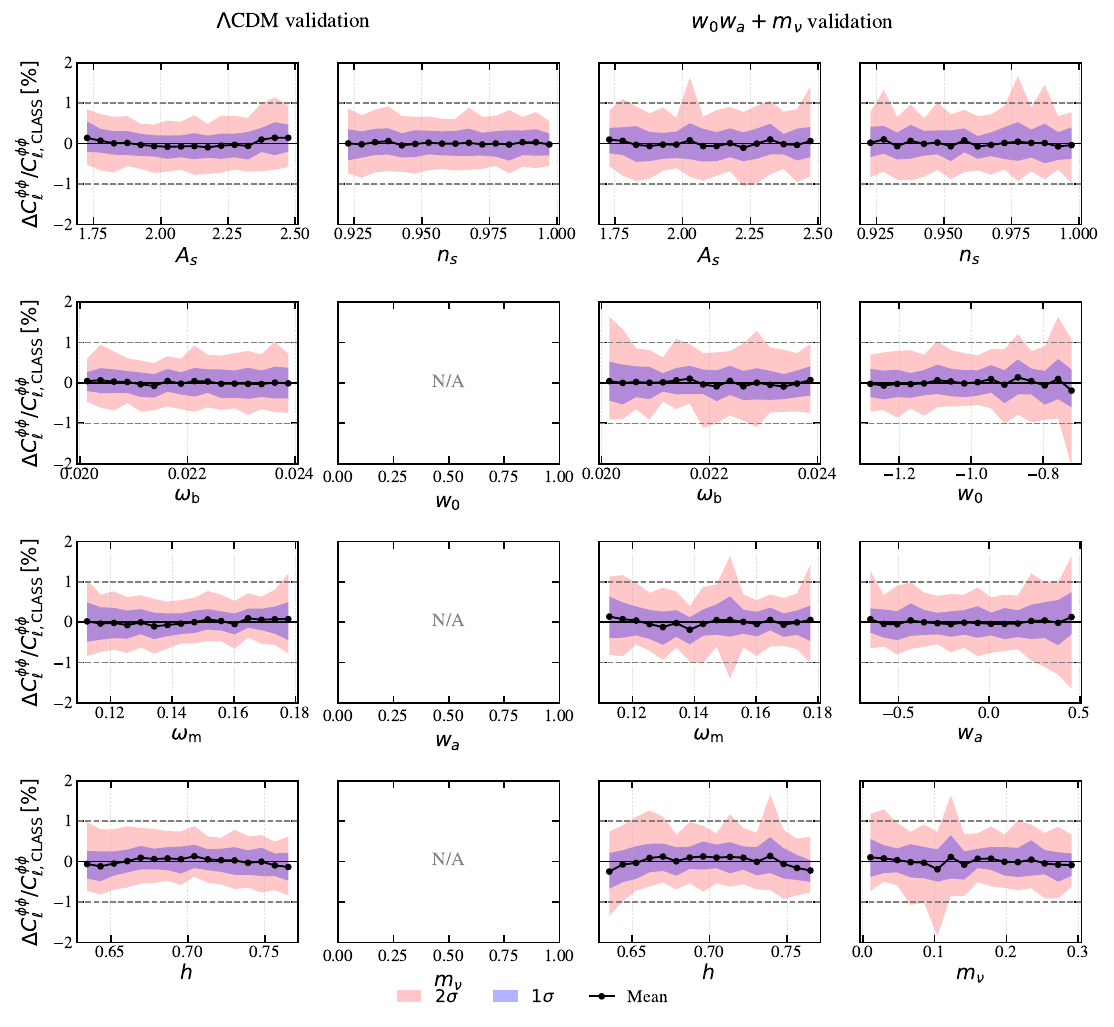}
    \caption{Emulator fractional-error distribution as a function of cosmological parameter value, averaged over \(\ell\) and over the remaining parameters, for \(\Lambda\)CDM cosmologies (left) and the full extended cosmology (right). The absence of significant monotonic trends or edge enhancements indicates that the emulator error is not concentrated in a particular region of the sampled parameter domain.}
    \label{fig:parameter_error}
\end{figure*}

We also considered the error distributions based on cosmological parameter values rather than the $\ell$ distribution alone to see if there is any bias that would affect inference. For each binned value of a parameter we average the relative error over the other parameters and $\ell$ and again look at the error distribution and its 1 and 2 sigma interval. Fig.~\ref{fig:parameter_error} shows no significant monotonic trends with cosmological parameter values and no clear enhancement of the error near the edges of the training domain.

\section{Performance}\label{sec:performance}
It is not straightforward to compare CMB emulators in a completely uniform way, since existing models differ in their training data, nonlinear prescriptions, output conventions, parameter ranges, and target spectra. A direct comparison on our validation set would therefore not be fully meaningful: our spectra are generated with the CLASS settings and HMcode 2020 prescription described above, while other emulators were trained on different Boltzmann-code versions, accuracy settings, or cosmological subspaces. We therefore focus here on the qualitative comparison most relevant for practical use: parameter coverage, portability, and integration into inference pipelines.

Concerning the runtime evaluation, a full standardization of the machine and environment on which we are computing would be necessary, but different environments and even hardware benefit different emulators, e.g. GPUs for NNs. 
Unlike neural networks, which often require specific Python environments (e.g., TensorFlow or PyTorch versions) and heavy libraries, a symbolic expression requires only standard mathematical functions, allowing for seamless integration into C++, Fortran, Python or other programming languages.

Compared to existing CMB emulators, CMBolic occupies a distinct niche. Direct neural-network emulators such as CosmoPower achieve high accuracy for CMB spectra, including lensing, and have been validated in end-to-end likelihood analyses, but are organised as separate emulator families for different cosmological extensions rather than as a single joint $w_0w_a+\Sigma m_\nu$ emulator. Modular approaches such as CLASSNET offer a different advantage, namely acceleration within the Boltzmann code itself, but their end-to-end speed-up for CMB harmonic spectra is more modest because the line-of-sight integration remains the dominant cost. At the other extreme, \texttt{capse.jl} demonstrates that neural-network emulators can reach extremely fast, differentiable spectrum evaluation. In this context, the main advantage of CMBolic is not maximal raw speed, but the combination of Stage-4-relevant precision, portability, and a single symbolic expression valid across the joint extended parameter space considered here. For practical MCMC applications, the evaluation time of CMBolic is already well below the likelihood overhead, so further reductions in spectrum-evaluation time would have limited impact on end-to-end runtime  (although this is not the only use case). 
For an isolated single-spectrum comparison on the same machine, CLASS takes a median of \(12.94\,{\rm s}\) to compute \(C_\ell^{\phi\phi}\) with the accuracy settings used for training. A precompiled C implementation of CMBolic, evaluated on the same fixed 5000-point multipole grid with all \(\ell\)-dependent quantities precomputed, takes \(6.3\times10^{-2}\,{\rm ms}\) per spectrum without fast-math compiler optimizations, corresponding to a single-spectrum speed-up of \(2.1\times10^5\). With aggressive fast-math optimization this decreases to \(2.1\times10^{-2}\,{\rm ms}\), corresponding to a speed-up of \(6.3\times10^5\). Since such microbenchmarks are sensitive to compiler flags, hardware, and output-grid conventions, our main performance claim is based on the end-to-end ACT DR6+Planck inference comparison.

A neat aspect of CMBolic is the compactness of its functional representation. Neural-network emulators can formally be regarded as analytic maps, but their explicit form would require all weights, biases, activation functions, input/output normalizations, and, for PCA-based variants, the reconstruction basis. For example, a CosmoPower-like PCA network with four 512-node hidden layers and a few hundred PCA coefficients already corresponds to $O(10^6)$ scalar constants once the PCA basis is included; even older or smaller multilayer-perceptron approaches such as CosmoNet are naturally at the $O(10^5)$ parameter level when written out explicitly. Modern neural-network emulators such as \texttt{capse.jl} are similarly compact as software objects, but not as literal closed-form expressions: their description length is set by the trained network weights, normalization arrays, output grids, and post-processing steps, typically implying at least $10^5$ to $10^6$ primitive constants for CMB-spectrum emulation. Compared to these estimates, CMBolic contains only 45 fitted scalar coefficients.

Further raw speed-ups are possible by batching evaluations or by implementing the symbolic expression in GPU-enabled frameworks such as PyTorch \citep{paszke2019pytorch}, but for the applications considered here the dominant benefit is the ability to perform fast, portable inference over a joint extended cosmological parameter space.

\subsection{Inference comparison}
For practical usage of this suite, an even better metric of functionality than relative error is the agreement of inference contours when using this emulator to contours obtained with full Boltzmann codes such as CLASS. 
It has recently been shown that symbolic emulators obtain unbiased constraints for Weak Lensing analyses \citep{Bartlett_2025}, so in this section we test whether the same is true for CMB lensing.
We implement our emulator as a theory model in Cobaya \citep{torrado2021cobaya}. We run an MCMC comparison of contours between CMBolic and CLASS using the ACT DR6 lensing likelihood applied to real ACT DR6 + Planck data, adopting the \texttt{actplanck\_extended} variant of the likelihood, which adds low multipoles from the Planck survey \citep{carron2022cmb} for more constraining power than the pure ACT option. We run 24 chains for both theory classes, we use the Gelman-Rubin criterion \citep{gelman1992inference} for convergence of $|R-1|<0.01$, a check of effective sample sizes over 400 for each parameter, as well as manual visual checks of the convergence of the chains. 

CMB lensing is usually not used for inference on its own, but is combined with the primary CMB temperature and polarization spectra or with other LSS probes, leading to much tighter parameter constraints. Although lensing-only data constrain all cosmological parameters to some degree, their strongest constraining power is on parameters controlling the late-time growth of structure. We therefore impose weak external information on parameters that are poorly constrained by lensing alone. Specifically, we use Planck-informed
Gaussian priors originally based on Big Bang Nucleosynthesis (BBN) priors on the physical baryon density,
\(\omega_{\rm b}=0.02237\pm0.00015\), and on the scalar spectral index,
\(n_{\rm s}=0.9649\pm0.0042\), matching the Planck 2018 baseline
\(\Lambda\)CDM constraints \citep{aghanim2020planck}.
Informative Gaussian priors on the scalar spectral index follow the treatment adopted in the Planck lensing analyses \citep{ade2016planck,aghanim2020planck}, where the lensing-only constraints were found to be only weakly sensitive to this choice. The remaining sampled parameters are assigned broad uniform priors: \(h\in[0.63,0.77]\), \(A_{\rm s}\in[1.7,2.5]\times10^{-9}\), \(\omega_{\rm m}\in[0.11,0.18]\), \(w_0\in[-1.3,-0.7]\), \(w_a\in[-0.7,0.5]\), and \(\Sigma m_\nu\in[0,0.3]\,{\rm eV}\).

\begin{figure*}
\includegraphics[width=\linewidth]{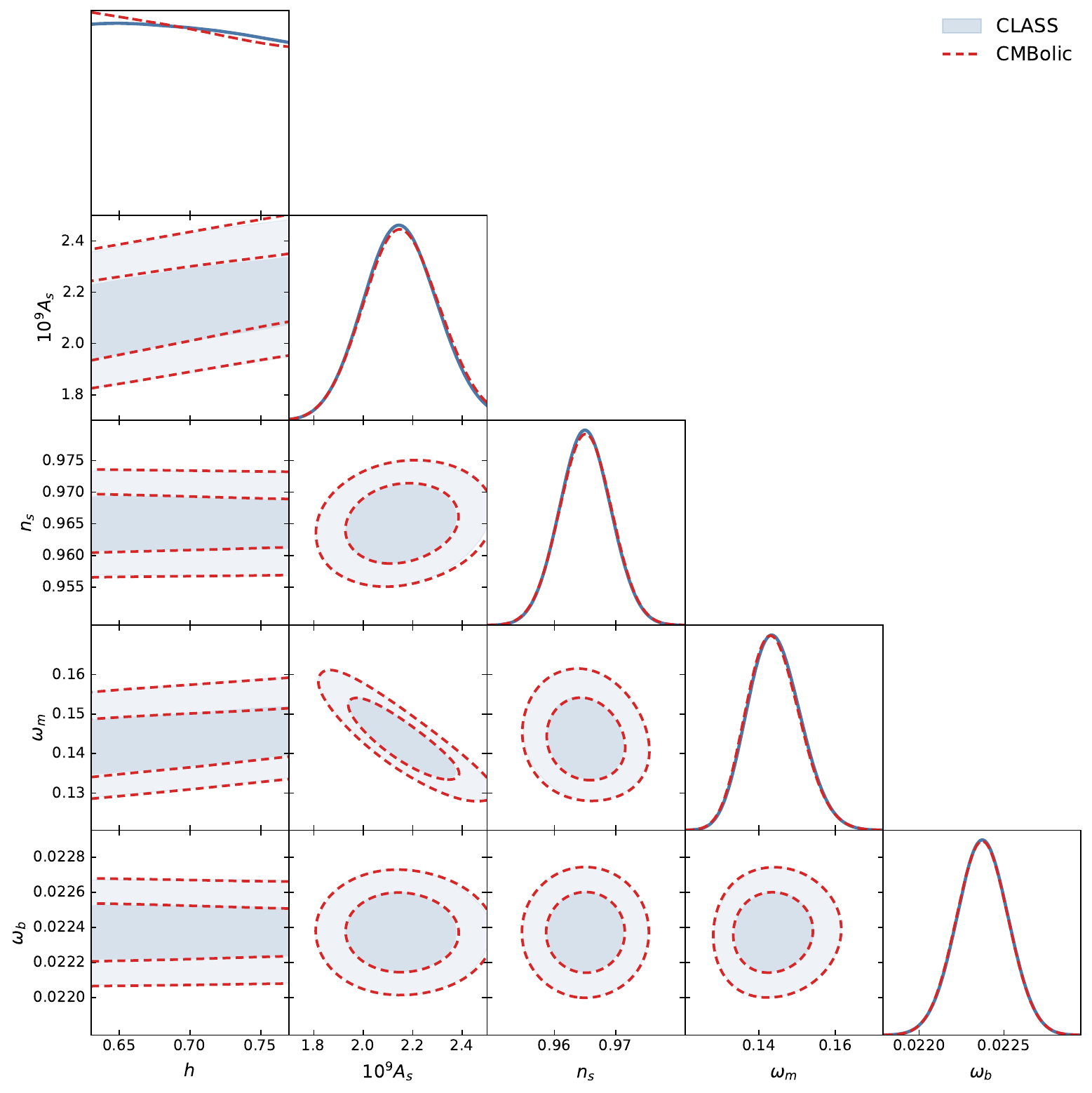}\caption{Comparison of posterior constraints in the $\Lambda$CDM parameter space obtained with CLASS (blue shaded contours) and CMBolic (red dashed contours), using the ACT DR6 + Planck lensing-only likelihood. The agreement is very good across both the marginalized one-dimensional posteriors and the two-dimensional parameter contours.}\label{fig:contour_lcdm}
\end{figure*}

\begin{figure*}
\includegraphics[width=\linewidth]{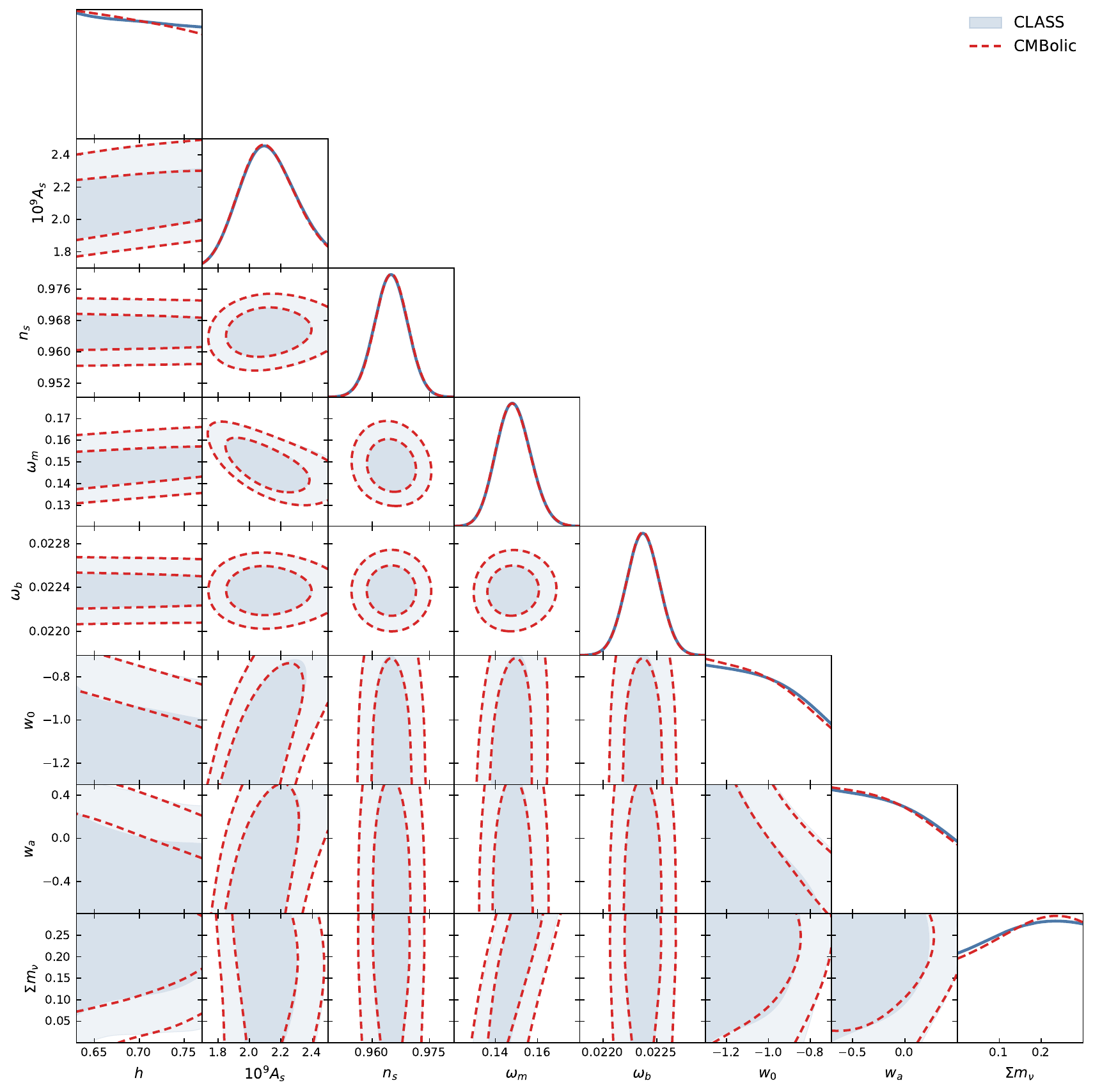}\caption{Comparison of posterior constraints in the extended cosmological parameter space obtained with CLASS (blue shaded contours) and CMBolic (red dashed contours), using the ACT DR6 + Planck lensing-only likelihood. The agreement is very good across both the marginalized one-dimensional posteriors and the two-dimensional parameter contours.}\label{fig:extended_contours}
\end{figure*}

Figs.~\ref{fig:contour_lcdm} and \ref{fig:extended_contours} compare the posterior constraints obtained with CLASS and CMBolic using the real
ACT DR6+Planck lensing-only likelihood. The contours and marginalized posteriors were produced with GetDist
\citep{Lewis:2019xzd}. The agreement is excellent across both the baseline \(\Lambda\)CDM and extended parameter spaces. To quantify the visual agreement, we compute the one-dimensional posterior overlap
\[\mathcal O_i =\int d\theta_i\,\min\left[p_{\rm CLASS}(\theta_i),p_{\rm CMBolic}(\theta_i)\right],\] using the normalized marginal densities for each sampled parameter. In the \(\Lambda\)CDM run, the smallest overlap is \(\mathcal O_i=0.984\), occurring for \(\omega_{\rm m}\). In the extended run, the smallest overlap is \(\mathcal O_i=0.983\), occurring for \(w_0\). Across all parameters and both inference tests, the largest posterior-mean shift is only \(0.038\,\sigma_{\rm CLASS}\). The neutrino-mass constraint is
also unchanged at the level relevant here, with 95\% upper limits \(\Sigma m_\nu<0.28596\,{\rm eV}\) for CLASS and \(\Sigma m_\nu<0.28623\,{\rm eV}\) for CMBolic. Thus the small visible differences in the triangle plots are negligible compared with the statistical
width of the lensing-only posterior.

In Fig.~\ref{fig:extended_contours} we compare the posterior constraints obtained with CLASS and CMBolic for the full ACT-Planck extended likelihood. This inference, which would take over two weeks to run with CLASS, took 164 seconds with CMBolic on the same 64-core machine. This corresponds to a wall-time speed-up of \(>7.4\times10^3\). 
As expected for a lensing-only analysis, the extended parameters are only weakly constrained, leading to broader marginalized posteriors than in the baseline \(\Lambda\)CDM run.

This comparison uses standard Cobaya MCMC and does not exploit the analytic differentiability of CMBolic. Gradient-based samplers could therefore provide additional acceleration in applications with differentiable likelihoods.

\subsection{Domain of validity and modelling choices}
\label{sec:domain}

The emulator is validated over the parameter ranges in
Table~\ref{table:parameter_ranges} and over the multipole interval
\(2\leq\ell\leq5500\). It should therefore not be used for precision inference
outside this domain without additional validation. Within the validated domain,
the formula should be interpreted as an emulator of the specific theoretical
prediction used to generate the training data: CLASS v3.3.0 with the ACT
DR6-inspired accuracy settings and the HMcode 2020 nonlinear prescription.

Not all modelling choices affect the prediction equally. Some choices, such as
the fixed value of \(\tau\), induce changes below the emulator validation error
for the lensing spectrum considered here. By contrast, changing the nonlinear
matter-power prescription changes the target function itself: for a Planck-like
fiducial cosmology, replacing HMcode 2020 with Halofit shifts
\(C_\ell^{\phi\phi}\) by up to \(4.4\%\), while using the linear matter power
spectrum instead of HMcode 2020 leads to much larger high-\(\ell\) differences.
Alternative nonlinear prescriptions, baryonic-feedback models, additional
cosmological parameters, or substantially broader parameter ranges should
therefore be validated separately, or incorporated through retraining or
dedicated correction factors, for which our emulator can still be a good symbolic basis.


\section{Conclusions}\label{sec:conclusions}
We have presented the first component of CMBolic, a symbolic emulator suite for CMB spectra, focusing here on the CMB lensing potential power spectrum \(C_\ell^{\phi\phi}\). The emulator is trained on high-accuracy CLASS spectra computed with HMcode 2020 and covers both \(\Lambda\)CDM cosmologies and an extended parameter space including the CPL dark-energy parameters \(w_0,w_a\) and the total neutrino mass \(\Sigma m_\nu\). To our knowledge, this is the first emulator for \(C_\ell^{\phi\phi}\) trained directly over the joint \(w_0w_a+\Sigma m_\nu\) extension of \(\Lambda\)CDM within a single model.

On independent validation spectra evaluated over \(2\leq\ell\leq5500\), CMBolic achieves mean absolute fractional errors of \(0.27\%\) in the \(\Lambda\)CDM subspace and \(0.32\%\) across the full extended parameter space. These residuals remain comfortably within the statistical error budget relevant for the CMB lensing applications considered here, including an optimistic Stage-4-like comparison, and show no significant systematic trends across the sampled cosmological parameter ranges. The emulator residuals are also much smaller than differences induced by common nonlinear matter-power modelling choices: for a Planck-like fiducial cosmology, the residual is well below the difference between HMcode 2020 and Halofit, and far below the discrepancy obtained when replacing HMcode 2020 by the linear matter power spectrum.

We further verified CMBolic at the inference level by replacing CLASS in Cobaya analyses of the real ACT DR6+Planck lensing-only likelihood. The resulting one- and two-dimensional posteriors agree closely with those obtained using CLASS in both the baseline \(\Lambda\)CDM and extended cosmological parameter spaces. The smallest one-dimensional posterior overlap is \(0.984\) in the \(\Lambda\)CDM run and \(0.983\) in the extended run, while the largest posterior-mean shift across all parameters is only \(0.038\sigma_{\rm CLASS}\). The neutrino-mass upper limit is likewise unchanged at the level relevant here. In the full extended inference test, replacing CLASS by CMBolic reduced the wall time from more than two weeks to \(164\) seconds on the same hardware, while preserving the posterior constraints.

The final emulator is a closed-form symbolic expression with only 45 fitted coefficients. It requires only elementary mathematical functions, is straightforward to implement in Python, C, Fortran, or other languages, and does not depend on external machine-learning libraries. This makes CMBolic complementary to neural-network emulators: its main advantage is not just raw speed, but the combination of portability, differentiability, compactness, and coverage of a joint extended cosmological parameter space. The present model should nevertheless be used within the parameter ranges and multipole interval on which it was trained and validated, and under the modelling assumptions described in Sect.~\ref{sec:domain}. Changes such as alternative nonlinear prescriptions, baryonic-feedback freedom, additional cosmological parameters, or substantially broader priors should be validated separately or incorporated through retraining or dedicated correction factors.

This is the first paper in the CMBolic series of CMB emulators. We leave the temperature and polarization anisotropy spectra for forthcoming work, where the lensing emulator developed here will be useful for modelling the lensing correction from unlensed to lensed CMB spectra. Alongside the \textsc{syren} family of symbolic emulators for matter power spectra, CMBolic is a step towards a complete symbolic-emulation toolkit for cosmological observables. We expect the present emulator to provide a fast, portable, and accurate way to compute CMB lensing spectra across an extended cosmological parameter space not previously covered by a single symbolic emulator.


\section*{Data Availability}

Symbolic emulators are available as a part of the CMBolic suite on GitHub in various languages for simple usage, as well as the implementation of our emulator as a theory class in Cobaya: \url{https://github.com/dvokrouhlicky/CMBolic}. Training datasets can be regenerated using the CLASS settings listed in Appendix~\ref{sec:appendix_A}.
\begin{acknowledgements}
      
We thank D. Farakou for early access to \texttt{Syrenclass} and many important improvements and suggestions, R. Calderon for advice on cosmological inference, B. Burlacu and G. Kronberger for help with the invaluable Operon code. 
This work was co-financed by the European Structural and Investment Funds and the Czech Ministry of Education, Youth and Sports (MSMT) of the Czech Republic 
 (Project FORTE – $\mathrm{PCZ.02.01.01/00/22\_008/0004632}$). 
 C.S. further acknowledges support by the Royal Society Wolfson Visiting Fellowship ``Testing the properties of dark matter with new statistical tools and cosmological data''.
D.J.B. acknowledges that support was provided by Schmidt Sciences, LLC.
H.D. is supported by Royal Society University Research Fellowship grant 211046.
\end{acknowledgements}

\bibliographystyle{aa} 
\bibliography{CMBolic_references}

\providecommand{\noopsort}[1]{}\providecommand{\singleletter}[1]{#1}%
\begin{thebibliography}{69}
\expandafter\ifx\csname natexlab\endcsname\relax\def\natexlab#1{#1}\fi

\bibitem[{Abazajian {et~al.}(2019)Abazajian, Addison, Adshead, Ahmed, Allen, Alonso, Alvarez, Anderson, Arnold, Baccigalupi, {et~al.}}]{abazajian2019cmb}
Abazajian, K., Addison, G., Adshead, P., {et~al.} 2019, arXiv preprint arXiv:1907.04473

\bibitem[{{AbdusSalam} {et~al.}(2025{\natexlab{a}}){AbdusSalam}, {Abel}, {Bartlett}, \& {Crispim Rom{\~a}o}}]{abdussalam2025royalsociety}
{AbdusSalam}, S., {Abel}, S., {Bartlett}, D., \& {Crispim Rom{\~a}o}, M. 2025{\natexlab{a}}, arXiv e-prints, arXiv:2510.20453

\bibitem[{{AbdusSalam} {et~al.}(2025{\natexlab{b}}){AbdusSalam}, {Abel}, \& {Rom{\~a}o}}]{abdussalam2025symbolic}
{AbdusSalam}, S., {Abel}, S., \& {Rom{\~a}o}, M.~C. 2025{\natexlab{b}}, \prd, 111, 015022

\bibitem[{Abitbol {et~al.}(2025)Abitbol, Abril-Cabezas, Adachi, Ade, Adler, Agrawal, Aguirre, Ahmed, Aiola, Alford, {et~al.}}]{abitbol2025simons}
Abitbol, M., Abril-Cabezas, I., Adachi, S., {et~al.} 2025, arXiv preprint arXiv:2503.00636

\bibitem[{Adame {et~al.}(2025)Adame, Aguilar, Ahlen, Alam, Alexander, Alvarez, Alves, Anand, Andrade, Armengaud, {et~al.}}]{adame2025desi}
Adame, A., Aguilar, J., Ahlen, S., {et~al.} 2025, Journal of Cosmology and Astroparticle Physics, 2025, 021

\bibitem[{Ade {et~al.}(2019)Ade, Aguirre, Ahmed, Aiola, Ali, Alonso, Alvarez, Arnold, Ashton, Austermann, {et~al.}}]{ade2019simons}
Ade, P., Aguirre, J., Ahmed, Z., {et~al.} 2019, Journal of Cosmology and Astroparticle Physics, 2019, 056

\bibitem[{Ade {et~al.}(2016)Ade, Aghanim, Arnaud, Ashdown, Aumont, Baccigalupi, Banday, Barreiro, Bartlett, Bartolo, {et~al.}}]{ade2016planck}
Ade, P.~A., Aghanim, N., Arnaud, M., {et~al.} 2016, Astronomy \& Astrophysics, 594, A15

\bibitem[{Aghanim {et~al.}(2020)Aghanim, Akrami, Ashdown, Aumont, Baccigalupi, Ballardini, Banday, Barreiro, Bartolo, Basak, {et~al.}}]{aghanim2020planck}
Aghanim, N., Akrami, Y., Ashdown, M., {et~al.} 2020, Astronomy \& Astrophysics, 641, A8

\bibitem[{Albers {et~al.}(2019)Albers, Fidler, Lesgourgues, Sch{\"o}neberg, \& Torrado}]{albers2019cosmicnet}
Albers, J., Fidler, C., Lesgourgues, J., Sch{\"o}neberg, N., \& Torrado, J. 2019, Journal of Cosmology and Astroparticle Physics, 2019, 028

\bibitem[{{Angulo} {et~al.}(2021){Angulo}, {Zennaro}, {Contreras}, {Aric{\`o}}, {Pellejero-Iba{\~n}ez}, \& {St{\"u}cker}}]{bacco2021nonlinear}
{Angulo}, R.~E., {Zennaro}, M., {Contreras}, S., {et~al.} 2021, \mnras, 507, 5869

\bibitem[{Archidiacono {et~al.}(2020)Archidiacono, Hannestad, \& Lesgourgues}]{archidiacono2020will}
Archidiacono, M., Hannestad, S., \& Lesgourgues, J. 2020, Journal of Cosmology and Astroparticle Physics, 2020, 021

\bibitem[{{Aric{\`o}} {et~al.}(2021){Aric{\`o}}, {Angulo}, \& {Zennaro}}]{bacco2021linear}
{Aric{\`o}}, G., {Angulo}, R.~E., \& {Zennaro}, M. 2021, arXiv e-prints, arXiv:2104.14568

\bibitem[{Asgari {et~al.}(2021)}]{KiDS:2020suj}
Asgari, M. {et~al.} 2021, Astron. Astrophys., 645, A104

\bibitem[{Bahl {et~al.}(2025)Bahl, Fuchs, Menen, \& Plehn}]{bahl2025mathcal}
Bahl, H., Fuchs, E., Menen, M., \& Plehn, T. 2025, arXiv preprint arXiv:2507.05858

\bibitem[{Balkenhol {et~al.}(2024)Balkenhol, Trendafilova, Benabed, \& Galli}]{balkenhol2024candl}
Balkenhol, L., Trendafilova, C., Benabed, K., \& Galli, S. 2024, Astronomy \& Astrophysics, 686, A10

\bibitem[{Bartlett {et~al.}(2023)Bartlett, Kammerer, Kronberger, Desmond, Ferreira, Wandelt, Burlacu, Alonso, \& Zennaro}]{bartlett2023precise}
Bartlett, D.~J., Kammerer, L., Kronberger, G., {et~al.} 2023, arXiv preprint arXiv:2311.15865

\bibitem[{{Bartlett} \& {Pandey}(2025)}]{Bartlett_2025}
{Bartlett}, D.~J. \& {Pandey}, S. 2025, arXiv e-prints, arXiv:2510.18749

\bibitem[{Bartlett {et~al.}(2024)Bartlett, Wandelt, Zennaro, Ferreira, \& Desmond}]{bartlett2024syren}
Bartlett, D.~J., Wandelt, B.~D., Zennaro, M., Ferreira, P.~G., \& Desmond, H. 2024, arXiv preprint arXiv:2402.17492

\bibitem[{Bolliet {et~al.}(2024)Bolliet, Spurio~Mancini, Hill, Madhavacheril, Jense, Calabrese, \& Dunkley}]{bolliet2024high}
Bolliet, B., Spurio~Mancini, A., Hill, J.~C., {et~al.} 2024, Monthly Notices of the Royal Astronomical Society, 531, 1351

\bibitem[{Bonici {et~al.}(2024)Bonici, Bianchini, \& Ruiz-Zapatero}]{Bonici2024Capse}
Bonici, M., Bianchini, F., \& Ruiz-Zapatero, J. 2024, The Open Journal of Astrophysics, 7

\bibitem[{Bonici {et~al.}(2025)Bonici, D'Amico, Bel, \& Carbone}]{Bonici:2025ltp}
Bonici, M., D'Amico, G., Bel, J., \& Carbone, C. 2025, JCAP, 2025, 044

\bibitem[{Burlacu {et~al.}(2020)Burlacu, Kronberger, \& Kommenda}]{10.1145/3377929.3398099}
Burlacu, B., Kronberger, G., \& Kommenda, M. 2020, in Proceedings of the 2020 Genetic and Evolutionary Computation Conference Companion, GECCO '20 (New York, NY, USA: Association for Computing Machinery), 1562–1570

\bibitem[{Calabrese {et~al.}(2025{\natexlab{a}})Calabrese, Hill, Jense, La~Posta, Abril-Cabezas, Addison, Ade, Aiola, Alford, Alonso, {et~al.}}]{calabrese2025atacama}
Calabrese, E., Hill, J.~C., Jense, H.~T., {et~al.} 2025{\natexlab{a}}, arXiv preprint arXiv:2503.14454

\bibitem[{Calabrese {et~al.}(2025{\natexlab{b}})}]{AtacamaCosmologyTelescope:2025nti}
Calabrese, E. {et~al.} 2025{\natexlab{b}}, JCAP, 11, 063

\bibitem[{Carron {et~al.}(2022)Carron, Mirmelstein, \& Lewis}]{carron2022cmb}
Carron, J., Mirmelstein, M., \& Lewis, A. 2022, Journal of Cosmology and Astroparticle Physics, 2022, 039

\bibitem[{Cava {et~al.}(2021)Cava, Orzechowski, Burlacu, de~Fran{\c{c}}a, Virgolin, Jin, Kommenda, \& Moore}]{DBLP:journals/corr/abs-2107-14351}
Cava, W. G.~L., Orzechowski, P., Burlacu, B., {et~al.} 2021, CoRR, abs/2107.14351 [\eprint{2107.14351}]

\bibitem[{Cranmer(2023)}]{cranmer2023interpretable}
Cranmer, M. 2023, arXiv preprint arXiv:2305.01582

\bibitem[{de~Fran{\c{c}}a \& Kronberger(2025)}]{de2025improving}
de~Fran{\c{c}}a, F.~O. \& Kronberger, G. 2025, arXiv preprint arXiv:2501.17848

\bibitem[{Farakou \& Skordis(2025)}]{Farakou:2025tuq}
Farakou, D. \& Skordis, C. 2025 [\eprint[arXiv]{2511.05093}]

\bibitem[{{Fendt} \& {Wandelt}(2007{\natexlab{a}})}]{Fendt_2007_Pico1}
{Fendt}, W.~A. \& {Wandelt}, B.~D. 2007{\natexlab{a}}, arXiv e-prints, arXiv:0712.0194

\bibitem[{{Fendt} \& {Wandelt}(2007{\natexlab{b}})}]{Fendt_2007_Pico2}
{Fendt}, W.~A. \& {Wandelt}, B.~D. 2007{\natexlab{b}}, \apj, 654, 2

\bibitem[{Gelman \& Rubin(1992)}]{gelman1992inference}
Gelman, A. \& Rubin, D.~B. 1992, Statistical science, 7, 457

\bibitem[{Grayeli {et~al.}(2024)Grayeli, Sehgal, Costilla~Reyes, Cranmer, \& Chaudhuri}]{grayeli2024symbolic}
Grayeli, A., Sehgal, A., Costilla~Reyes, O., Cranmer, M., \& Chaudhuri, S. 2024, Advances in Neural Information Processing Systems, 37, 44678

\bibitem[{Green \& Meyers(2025)}]{green2025cosmological}
Green, D. \& Meyers, J. 2025, Physical Review D, 111, 083507

\bibitem[{G{\"u}nther(2023)}]{gunther2023uncertainty}
G{\"u}nther, S. 2023, arXiv preprint arXiv:2307.01138

\bibitem[{G{\"u}nther {et~al.}(2022)G{\"u}nther, Lesgourgues, Samaras, Sch{\"o}neberg, Stadtmann, Fidler, \& Torrado}]{gunther2022cosmicnet}
G{\"u}nther, S., Lesgourgues, J., Samaras, G., {et~al.} 2022, Journal of Cosmology and Astroparticle Physics, 2022, 035

\bibitem[{Henderson {et~al.}(2016)Henderson, Allison, Austermann, Baildon, Battaglia, Beall, Becker, De~Bernardis, Bond, Calabrese, {et~al.}}]{henderson2016advanced}
Henderson, S., Allison, R., Austermann, J., {et~al.} 2016, Journal of Low Temperature Physics, 184, 772

\bibitem[{Hu \& Sugiyama(1994)}]{hu1994anisotropies}
Hu, W. \& Sugiyama, N. 1994, arXiv preprint astro-ph/9407093

\bibitem[{Kammerer {et~al.}(2025)Kammerer, Bartlett, Kronberger, Desmond, \& Ferreira}]{Kammerer:2025dbi}
Kammerer, L., Bartlett, D.~J., Kronberger, G., Desmond, H., \& Ferreira, P.~G. 2025, Astron. Astrophys., 701, A284

\bibitem[{Kaplinghat {et~al.}(2002)Kaplinghat, Knox, \& Skordis}]{Kaplinghat:2002mh}
Kaplinghat, M., Knox, L., \& Skordis, C. 2002, Astrophys. J., 578, 665

\bibitem[{Knabenhans {et~al.}(2021)}]{Euclid:2020rfv}
Knabenhans, M. {et~al.} 2021, Mon. Not. Roy. Astron. Soc., 505, 2840

\bibitem[{Knox(1995)}]{knox1995determination}
Knox, L. 1995, Physical Review D, 52, 4307

\bibitem[{Koza(1992)}]{10.5555/138936}
Koza, J.~R. 1992, Genetic programming: on the programming of computers by means of natural selection (Cambridge, MA, USA: MIT Press)

\bibitem[{Kronberger {et~al.}(2024)Kronberger, Burlacu, Kommenda, Winkler, \& Affenzeller}]{Kronberger_2024}
Kronberger, G., Burlacu, B., Kommenda, M., Winkler, S.~M., \& Affenzeller, M. 2024, Symbolic Regression (Chapman \& Hall / CRC Press)

\bibitem[{Lesgourgues(2011)}]{lesgourgues2011cosmic}
Lesgourgues, J. 2011, arXiv preprint arXiv:1104.2932

\bibitem[{Lewis(2025)}]{Lewis:2019xzd}
Lewis, A. 2025, JCAP, 08, 025

\bibitem[{Lewis \& Challinor(2011)}]{lewis2011camb}
Lewis, A. \& Challinor, A. 2011, Astrophysics source code library, ascl

\bibitem[{Limber(1953)}]{limber1953analysis}
Limber, D.~N. 1953, Astrophysical Journal, vol. 117, p. 134, 117, 134

\bibitem[{Lodha {et~al.}(2025)Lodha, Calderon, Matthewson, Shafieloo, Ishak, Pan, Garcia-Quintero, Huterer, Valogiannis, Ure{\~n}a-L{\'o}pez, {et~al.}}]{lodha2025extended}
Lodha, K., Calderon, R., Matthewson, W., {et~al.} 2025, arXiv preprint arXiv:2503.14743

\bibitem[{Madhavacheril {et~al.}(2024)}]{ACT:2023kun}
Madhavacheril, M.~S. {et~al.} 2024, Astrophys. J., 962, 113

\bibitem[{Mead {et~al.}(2021)Mead, Brieden, Tr{\"o}ster, \& Heymans}]{mead2021hmcode}
Mead, A., Brieden, S., Tr{\"o}ster, T., \& Heymans, C. 2021, Monthly Notices of the Royal Astronomical Society, 502, 1401

\bibitem[{Mead {et~al.}(2016)Mead, Heymans, Lombriser, Peacock, Steele, \& Winther}]{mead2016accurate}
Mead, A., Heymans, C., Lombriser, L., {et~al.} 2016, Monthly Notices of the Royal Astronomical Society, 459, 1468

\bibitem[{Mead {et~al.}(2015)Mead, Peacock, Heymans, Joudaki, \& Heavens}]{mead2015accurate}
Mead, A.~J., Peacock, J.~A., Heymans, C., Joudaki, S., \& Heavens, A.~F. 2015, Monthly Notices of the Royal Astronomical Society, 454, 1958

\bibitem[{Muthyala {et~al.}(2025)Muthyala, Sorourifar, Peng, \& Paulson}]{muthyala2025symantic}
Muthyala, M.~R., Sorourifar, F., Peng, Y., \& Paulson, J.~A. 2025, Industrial \& Engineering Chemistry Research

\bibitem[{Nygaard {et~al.}(2023)Nygaard, Holm, Hannestad, \& Tram}]{nygaard2023connect}
Nygaard, A., Holm, E.~B., Hannestad, S., \& Tram, T. 2023, Journal of Cosmology and Astroparticle Physics, 2023, 025

\bibitem[{Paszke {et~al.}(2019)Paszke, Gross, Massa, Lerer, Bradbury, Chanan, Killeen, Lin, Gimelshein, Antiga, Desmaison, Köpf, Yang, DeVito, Raison, Tejani, Chilamkurthy, Steiner, Fang, Bai, \& Chintala}]{paszke2019pytorch}
Paszke, A., Gross, S., Massa, F., {et~al.} 2019, PyTorch: An Imperative Style, High-Performance Deep Learning Library

\bibitem[{Qu {et~al.}(2024)}]{ACT:2023dou}
Qu, F.~J. {et~al.} 2024, Astrophys. J., 962, 112

\bibitem[{Radwan {et~al.}(2024)Radwan, Kronberger, \& Winkler}]{radwan2024comparison}
Radwan, Y.~A., Kronberger, G., \& Winkler, S. 2024, arXiv preprint arXiv:2406.03585

\bibitem[{Secco {et~al.}(2022)}]{DES:2022oqz}
Secco, L.~F. {et~al.} 2022, Phys. Rev. D, 105, 103537

\bibitem[{Sehgal {et~al.}(2019)Sehgal, Aiola, Akrami, Basu, Boylan-Kolchin, Bryan, Clesse, Cyr-Racine, Di~Mascolo, Dicker, {et~al.}}]{sehgal2019cmb}
Sehgal, N., Aiola, S., Akrami, Y., {et~al.} 2019, arXiv preprint arXiv:1906.10134

\bibitem[{Smith {et~al.}(2003)Smith, Peacock, Jenkins, White, Frenk, Pearce, Thomas, Efstathiou, \& Couchman}]{smith2003stable}
Smith, R.~E., Peacock, J.~A., Jenkins, A., {et~al.} 2003, Monthly Notices of the Royal Astronomical Society, 341, 1311

\bibitem[{Spurio~Mancini {et~al.}(2022)Spurio~Mancini, Piras, Alsing, Joachimi, \& Hobson}]{spurio2022cosmopower}
Spurio~Mancini, A., Piras, D., Alsing, J., Joachimi, B., \& Hobson, M.~P. 2022, Monthly Notices of the Royal Astronomical Society, 511, 1771

\bibitem[{Sui {et~al.}(2025)Sui, Bartlett, Pandey, Desmond, Ferreira, \& Wandelt}]{Sui:2024wob}
Sui, C., Bartlett, D.~J., Pandey, S., {et~al.} 2025, Astron. Astrophys., 698, A1

\bibitem[{Takahashi {et~al.}(2012)Takahashi, Sato, Nishimichi, Taruya, \& Oguri}]{takahashi2012revising}
Takahashi, R., Sato, M., Nishimichi, T., Taruya, A., \& Oguri, M. 2012, The Astrophysical Journal, 761, 152

\bibitem[{Thing \& Koksbang(2025)}]{Thing_2025}
Thing, M. \& Koksbang, S. 2025, Journal of Cosmology and Astroparticle Physics, 2025, 040

\bibitem[{To {et~al.}(2023)To, Rozo, Krause, Wu, Wechsler, \& Salcedo}]{to2023linna}
To, C.-H., Rozo, E., Krause, E., {et~al.} 2023, Journal of Cosmology and Astroparticle Physics, 2023, 016

\bibitem[{Torrado \& Lewis(2021)}]{torrado2021cobaya}
Torrado, J. \& Lewis, A. 2021, Journal of Cosmology and Astroparticle Physics, 2021, 057

\bibitem[{Virgolin \& Pissis(2022)}]{virgolin2022symbolic}
Virgolin, M. \& Pissis, S.~P. 2022, Transactions on Machine Learning Research [\eprint{arXiv:2207.01018}]

\bibitem[{Vokrouhlicky \& Skordis(2026)}]{vokrouhlicky2026}
Vokrouhlicky, D. \& Skordis, C. 2026, in preparation

\end{thebibliography}

\begin{appendix}

\section{CLASS settings}\label{sec:appendix_A}
The full recommended accuracy settings for CLASS from the ACT DR6 analysis are
\begin{verbatim}
"N_ncdm": 1,
"N_ur": 2.0308,
"T_cmb": 2.7255,
"YHe": 'BBN',
"non_linear": 'hmcode',
"hmcode_version": '2020',
"recombination": 'HyRec',
"lensing": 'yes',
"output": 'lCl , tCl , pCl , mPk',
"modes": 's',
"l_max_scalars": 9500,
"delta_l_max": 1800,
"P_k_max_h/Mpc": 100.,
"l_logstep": 1.025,
"l_linstep": 20,
"perturbations_sampling_stepsize": 0.05,
"l_switch_limber": 30.,
"hyper_sampling_flat": 32.,
"l_max_g": 40,
"l_max_ur": 35,
"l_max_pol_g": 60,
"ur_fluid_approximation": 2,
"ur_fluid_trigger_tau_over_tau_k": 130.,
"radiation_streaming_approximation": 2,
"radiation_streaming_trigger_tau_over_tau_k": 240.,
"hyper_flat_approximation_nu": 7000.,
"transfer_neglect_delta_k_S_t0": 0.17,
"transfer_neglect_delta_k_S_t1": 0.05,
"transfer_neglect_delta_k_S_t2": 0.17,
"transfer_neglect_delta_k_S_e": 0.17,
"accurate_lensing": 1,
"start_small_k_at_tau_c_over_tau_h": 0.0004,
"start_large_k_at_tau_h_over_tau_k": 0.05,
"tight_coupling_trigger_tau_c_over_tau_h": 0.005,
"tight_coupling_trigger_tau_c_over_tau_k": 0.008,
"start_sources_at_tau_c_over_tau_h": 0.006,
"l_max_ncdm": 30,
"tol_ncdm_synchronous": 1e-6,
"Omega_Lambda": 0,
"fluid_equation_of_state": 'CLP'
\end{verbatim}

\end{appendix}
\end{document}